\documentclass[twocolumn]{article}
\usepackage[T1]{fontenc}
\usepackage[numbers,sort&compress]{natbib}
\usepackage{amsmath, fullpage}
\usepackage{authblk}
\usepackage{graphicx}
\usepackage{comment}
\usepackage{amsmath}
\usepackage{amssymb}
\usepackage[font=small, labelfont=bf]{caption}
\usepackage{subcaption}
\usepackage[dvipsnames]{xcolor}
\usepackage[normalem]{ulem}
\usepackage{dblfloatfix} 
\usepackage{abstract}
\usepackage[version=4]{mhchem} 
\usepackage{subcaption}
\usepackage{graphicx}
\usepackage{xcolor}
\usepackage{soul}
\usepackage{siunitx}
\usepackage{ulem} 

\begin{document}

\title{Efficient formation and identification of  single emitters in 4H-SiC following maskless heavy ion implantation}

\author[1]{S. Eserin} 
\author[2,3]{L. K. S. Zimmermann} 
\author[1,4]{E. Schneider} 
\author[1,4]{M. Ludlow} 
\author[2,3]{J. Heiler} 
\author[1,4]{O. G. Lloyd-Willard} 
\author[5]{K. Stockbridge} 
\author[5]{G. Aresta} 
\author[1,4]{L. Antwis} 
\author[1,4]{S. K. Clowes} 
\author[6]{M. Hofmann} 
\author[2]{S. Kucera} 
\author[6]{P. Berwian} 
\author[2,3]{F. Kaiser} 
\author[1]{B. N. Murdin} 

\affil[1]{Advanced Technology Institute, University of Surrey, Guildford, GU2 7XH, UK}
\affil[2]{Quantum Materials, Luxembourg Institute of Science and Technology (LIST), 28 Avenue des Hauts Fourneaux, Belval, 4362 Esch-sur-Alzette, Luxembourg}
\affil[3]{ Department of Physics and Materials Science, University of Luxembourg, 2 Avenue de l’Université, Belval, 4365 Esch-sur-Alzette, Luxembourg}
\affil[4]{Surrey Ion Beam Centre, University of Surrey, Guildford, GU2 7XH, UK}
\affil[5]{Ionoptika Ltd, Eastleigh, SO53 4BZ, UK}
\affil[6]{Fraunhofer Institute for Integrated Systems and Device Technology IISB, Schottkystraße 10, 91058 Erlangen, Germany}

\twocolumn[
    \maketitle
    \begin{abstract}
 
Single photon emitters in silicon carbide (SiC) are a leading platform for scalable quantum technologies. Recent interest has focused on oxygen-vacancy-related emitters, which show exceptionally high optical brightness and strong spin readout contrast. One barrier to scalable quantum devices based on these emitters is the challenge of maskless formation and rapid identification. 
Here, we demonstrate the formation of isolated bright single emitters in 4H-SiC, using low-energy maskless implantation of heavy ions bismuth and tin. Following annealing, up to \SI{18}{\percent} of implanted sites host a single emitter, with optimal yields achieved at annealing temperatures of $900\text{--}1000$\si{\degreeCelsius}. Occupancy statistics are modelled to estimate the implantation dose that maximises single-emitter yield. We introduce a tiered characterisation scheme, where a simple intensity threshold isolates single-emitter candidates, confirmed through photon correlation measurements, after which correlations between polarisation, saturation count rate and magnetic resonance frequency assign emitter type. It is shown that time-consuming low-temperature spectroscopy is not necessary to distinguish emitter types. Together, maskless heavy-ion implantation and selective screening offer an efficient route to forming and rapidly identifying near-surface single emitters for room-temperature quantum technologies such as quantum sensing. 

    \end{abstract}
    \vspace{1em}
]

\section*{Introduction} 

Silicon carbide (SiC) has emerged as a leading semiconductor platform for quantum technologies, combining wafer-scale industrial processing with a variety of quantum emitters. SiC is, therefore, a compelling route towards scalable quantum devices integrated with nanophotonic and electronic structures. 

In 4H-SiC, the silicon vacancy~V$_\text{Si}$ is amongst the most studied emitters, notably due to its high-sensitivity quantum sensing \cite{ohshima2018}, as well as its successful integration into nanophotonic and electronic structures \cite{babin2022, Lukin2023}.  The divacancies ~V$_\text{Si}$V$_\text{C}$ (PL1-4) have also demonstrated coherent spin–photon interfacing and control of coupled nuclear-spin memories, making both emitter families suitable for quantum communication, computation and device integration \cite{he2024, nagy2019}.

Beyond these, the PL5\text{--}8 centres, previously referred to as "modified divacancies" \cite{he2024}, have recently been reassigned to oxygen-vacancy complexes O$_\text{C}$V$_\text{Si}$ \cite{hu2026, chen2026}, in agreement with theoretical calculations \cite{kobayashi2023, zhao2026}. Within this family, the PL6 is particularly promising as it closely resembles the NV$^{-}$ centre in diamond, combining long spin coherence, near-infrared emission within the second biological window \cite{li2025}, and high optically detected magnetic resonance (ODMR) spin contrast \cite{li2022}, all accessible at room temperature. Here, we demonstrate the formation of PL4\text{--}6 centres using heavy ion implantation in 4H-SiC, and develop a tiered characterisation scheme that allows for the rapid identification of promising emitters without the need for cryogenic systems.

Emitters can be generated using different approaches and are either formed from vacancies in the host lattice, as for the silicon vacancy V$_\text{Si}$, from an incorporated foreign species, or from a combination of the two, as for the oxygen vacancy. Electron, neutron, and proton irradiation \cite{babin2022, kasper2020, ohshima2018}, ion implantation \cite{babin2022, he2023, he2024, hu2026} and laser writing \cite{hao2025, feije2026} are the most common techniques employed for emitter creation.
Focused ion beam (FIB) implantation is of particular interest for maskless generation of emitters with deterministic spatial control, placing emitters precisely within nanophotonic structures such as waveguides \cite{babin2022} and pin diode structures to allow post-fabrication tuning of spin-optical properties \cite{scheller2025, zeledon2026}. FIB implantation also offers the most flexibility in the choice of emitter type, since the accessible ion species are limited only by the ion source  \cite{bischoff2016, hoeflich2023}. 

Heavy-ion implantation is known to generate intrinsic emitters in 4H-SiC, where the implanted species is not itself optically active, but instead produces a high density of local vacancies from which vacancy-based complexes form upon annealing \cite{kobayashi2022}. The formation of oxygen-containing PL5 and PL6 centres in this study is attributed to surface-bound oxygen or oxygen incorporated into the lattice during growth, which subsequently forms oxygen-vacancy centres upon annealing. A subsequent thermal anneal then serves to recover lattice damage and provide the thermal energy necessary for the vacancies to migrate and form the desired optically active complex \cite{karsthof2020, lee2021}. Previous heavy ion studies have relied on high-energy, broad-area implantation and ensemble measurements \cite{kobayashi2022}. Its use in maskless placement and formation of isolated single emitters has not been systematically explored. However, focused heavy-ion beam implantation targeting the generation of single emitters has been used to demonstrate the formation of tin-vacancy centres in diamond \cite{cheng2025}.

In this work, we fabricate emitters in 4H-SiC by focused heavy-ion implantation of tin (Sn) and bismuth (Bi) followed by subsequent annealing. This provides insight into the influence of heavy-ion implantation on the annealing dynamics of the emitter. Confocal photoluminescence mapping combined with second-order autocorrelation measurements was used to assess and model the generation of isolated single emitters. ODMR, low-temperature photoluminescence (LTPL), polarisation (POL) and saturation (SAT) measurements are then combined to classify the emitters by emitter type. Correlations between these metrics are used to define an efficient scheme targeting the rapid identification of PL6 centres for scalable quantum technologies.

\begin{figure*}[ht!]
\centering

        \includegraphics[width=\linewidth]{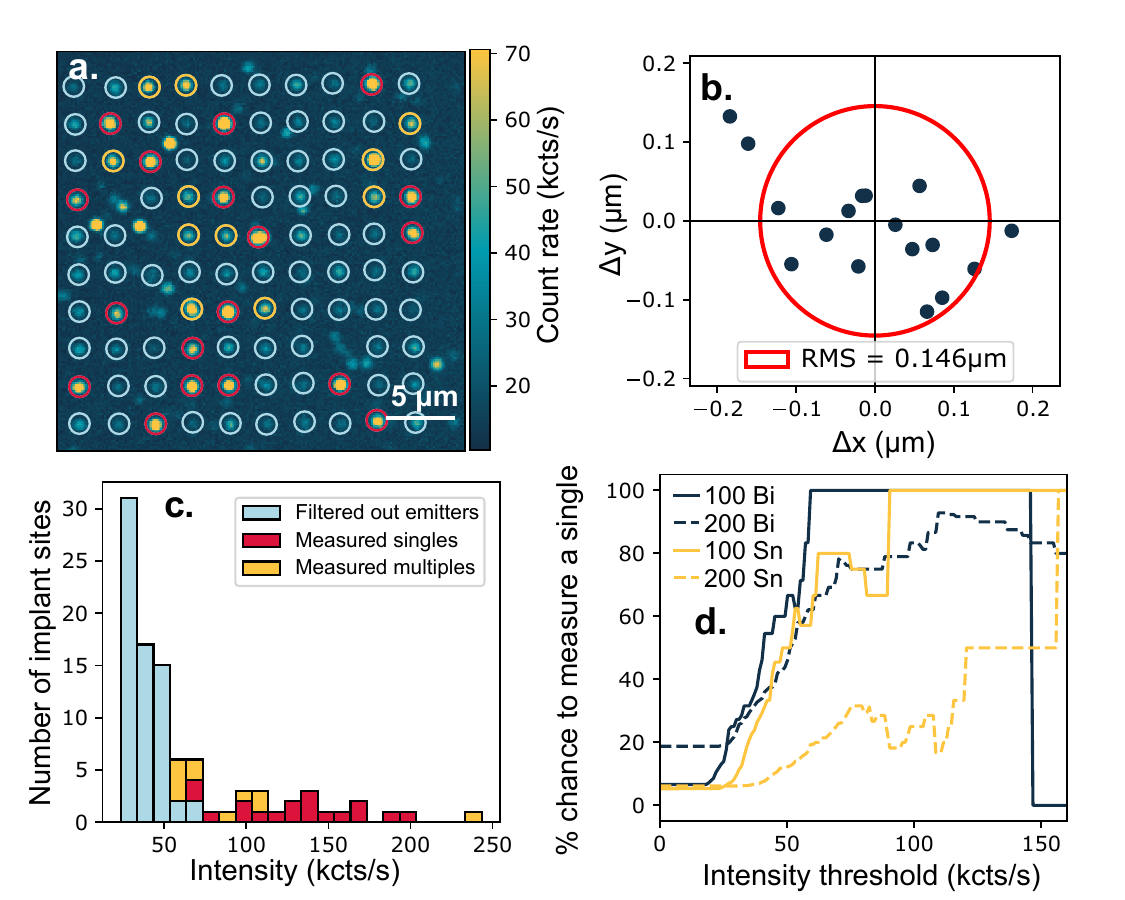}

    \caption{\textbf{a.} Confocal photoluminescence map of a Bi$^{2+}$ array, implanted with \SI{200}{ions/spot} and annealed at \SI{900}{\degreeCelsius}, under \SI{5}{mW} of \SI{940}{nm} off-resonant excitation. Red circles indicate confirmed single emitters, yellow circles indicate multiple emitters, and light blue circles are filtered-out implants due to a combination of count rate/spot shape.  \textbf{b.} Deviations of the fitted single-emitter positions from the idealised grid for the array in  \textbf{(a)}, showing an RMS displacement of \SI{146}{nm}. \textbf{c.} Histogram of intensities for all fitted emitter locations for the array in  \textbf{(a)}. \textbf{d.} Percentage change to measure a single emitter as a function of intensity threshold, shown here for Bi$^{2+}$ and Sn$^{2+}$ (blue and yellow, respectively), and for 100 and \SI{200}{ions/spot} (solid/dashed lines, respectively), for the \SI{900}{\degreeCelsius} annealed sample. The dashed blue line corresponds to the histogram in panel c.}
    \label{fig:pl_map_and_hbt}
\end{figure*}

\section*{Results} 

Nine SiC samples were implanted with Bi$^{+/2+}$ and Sn$^{+/2+}$ ions using a maskless FIB technique (Ionoptika QOne) at the UK National Ion Beam Centre. Each array features $10\times10$ spots of a chosen species and dose. The ion beam spot size is estimated from the FIB imaging mode resolution of \SI{14}{\nano\meter}, this sets a lower bound on the implant accuracy. The dose was controlled by pulsing the beam of known current with an electrostatic blanker. The FIB tool used is capable of precise counting of individual ion impact events \cite{schneider2026}, but for the larger doses used here, this function was not implemented. Ions were accelerated with a fixed \SI{25}{kV} anode voltage, giving \SI{25}{keV} and \SI{50}{keV} energy for singly and doubly charged species, respectively. Eight samples were annealed at 600\text{--}\SI{1100}{\degreeCelsius} and the ninth was left as an unannealed reference sample. Further details can be found in the methods.

Photoluminescence (PL) mapping was performed in a home-built scanning confocal microscope designed to suppress silicon V$_\text{Si}^-$ emission. 

An example PL map is shown in Fig. \ref{fig:pl_map_and_hbt}a for an array implanted with 200~Bi$^{2+}$ ions/spot and annealed at \SI{900}{\degreeCelsius}. The emitters that reside on the implantation grid are circled. Emitters that lie outside the implantation grid were excluded from further investigation as they presumably stem from naturally occurring emitter rather than from the implantation. Additionally, spots with a distorted shape are considered unsuccessful, as these indicate multiple emitters, either implantation-induced or in combination with a naturally occurring emitter. Unlike lighter ions (e.g. He/N/C), heavy ions create dense collision cascades, which can lead to higher vacancy concentrations in a single implant spot. These naturally high levels of residual damage are attributed to the fluorescence observed in the 'filtered-out' locations.

\begin{figure*}[hb!]

        \includegraphics[width=\linewidth]{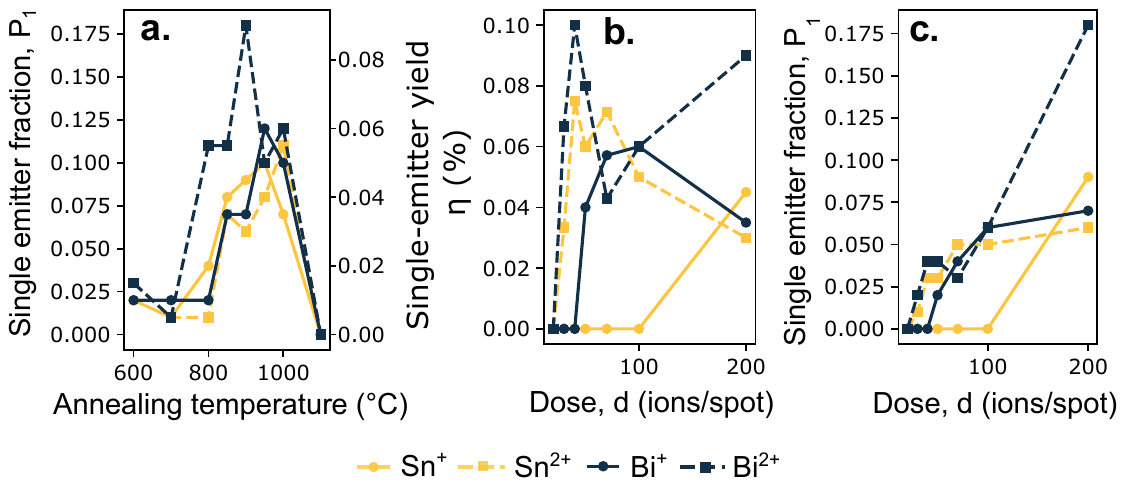}
    
    \caption{\textbf{a.} Single-emitter fraction and single-emitter yield as a function of annealing temperature for the Bi$^{+/2+}$, and Sn$^{+/2+}$  \SI{200}{ions/spot} arrays. \textbf{b.} The single-emitter yield as a function of the dose for each species for the \SI{900}{\degreeCelsius} sample. \textbf{c.} The single-emitter fraction for the same species and doses annealed at \SI{900}{\degreeCelsius}.}
    \label{fig:emitter_yields}
\end{figure*}

Implantation-position accuracy was extracted by comparing the fitted PL spot centre with an idealised grid \cite{babin2022}. Fig. \ref{fig:pl_map_and_hbt}b shows the scatter of these deviations from which a root mean square (RMS) displacement of \SI{146}{\nano\meter} is obtained for Bi$^{2+}$. Details of the grid fitting and RMS displacement calculation are given in the supplementary materials. The RMS displacement includes a contribution from the ion straggle. The straggle for Bi$^{2+}$ at \SI{25}{kV} anode voltage is \SI{3.7}{\nano\meter} (see supplementary materials), and is therefore negligible, so the scatter in implant position dominates the displacement. This technique is similar to that developed previously \cite{babin2022} where implantation through a \SI{50}{\nano\meter} lithographically defined mask resulted in a measured RMS spread in emitter position of \SI{53.6}{\nano\meter} and was therefore dominated by the hole size. Historically, the ion beam diameter produced from the resolution in imaging mode is used to obtain the beam diameter, but in this case it was of order 14~nm, significantly smaller than the spread in emitter position. This indicates that the imaging resolution is not a good measure of implant accuracy and that additional factors such as an error in the scanning of the beam or broad beam tails could influence the final scatter in implanted emitter positions \cite{Masteghin2026}.

Following the PL mapping, spots on the grid were selected for further analysis based on their brightness. A fixed count rate selection threshold of $>$40\,kcts/s at \SI{5}{\milli\watt} excitation power was applied to strongly suppress residual damage and V$_\text{Si}^{-}$ emission, while the brighter PL5-8 centres are expected to lie above this threshold \cite{hu2026}.  Spots below the threshold were marked with a blue circle on Fig. \ref{fig:pl_map_and_hbt}a and treated as unsuccessful implants. The remaining spots were characterised with the second-order autocorrelation function $g^{(2)}(\tau)$ using a Hanbury-Brown-Twiss setup (see methods). Those with a $g^{(2)}(0)<0.5$ were classified as single emitters and marked red, and the rest, containing multiple emitters, were marked yellow. All classified spots were then binned by their measured count rate. The resulting histogram is shown for the Bi$^{2+}$ \SI{200}{ions/spot} array in Fig. \ref{fig:pl_map_and_hbt}c. From this distribution,  the survival probability of single-emitters with a count-rate threshold can be calculated (Fig. \ref{fig:pl_map_and_hbt}d). This figure shows how process yield can be increased: raising the threshold reduces the number of successful single emitters that can be obtained, but increases the likelihood that selected emitters are singles.

\begin{figure*}[ht!]
        \includegraphics[width=\textwidth]{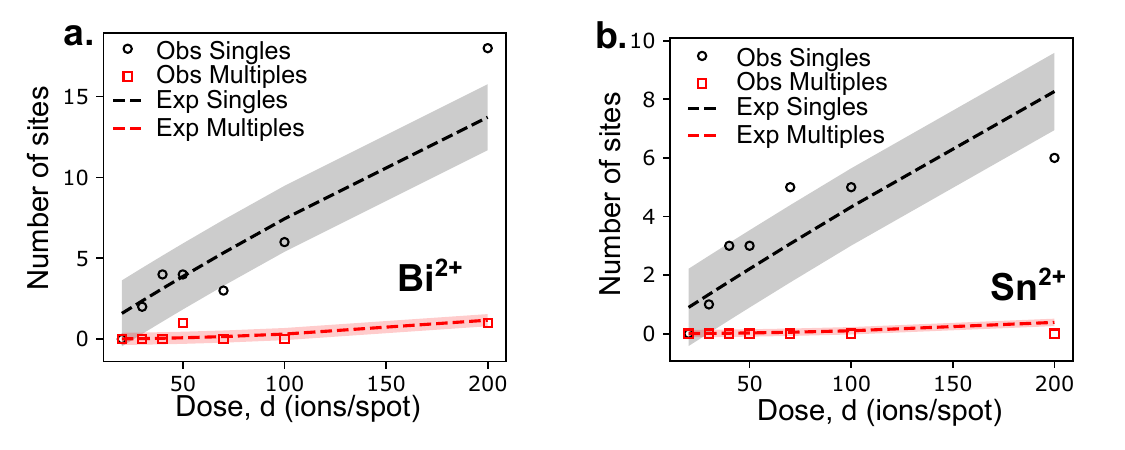}

    \caption{Numbers of single and multiple emitters as a function of dose for \textbf{a.} Bi$^{2+}$ and \textbf{b.} Sn$^{2+}$, with the expected number of sites given by a Poisson model in the low-dose regime shown as dashed lines with best-fit $\hat\alpha$ of Bi$^{2+}$: (8.1$\pm$1.3)$\times$10$^{-4}$ emitters per ion and Sn$^{2+}$: (4.5$\pm$0.9)$\times$10$^{-4}$ emitters per ion. The shaded bands in each plot represent the standard deviation of the residuals from each fit.}
    \label{fig:Poisson_analysis}
\end{figure*}

\subsection*{Optimal dose estimation and formation efficiency}

To estimate the optimal dose and formation efficiency of emitters, we apply the described detection of single emitters to all doses and annealing temperatures and analyse the data by calculating two metrics: the single-emitter fraction $p_{1}$, and the single-emitter yield, $\eta$.  The single-emitter fraction is the fraction of implanted spots hosting exactly one emitter, while the single-emitter yield is the number of single emitters produced per implanted ion.  The variation in single-emitter fraction and single-emitter yield as a function of annealing temperature and dose is shown in Fig. \ref{fig:emitter_yields}. These demonstrate an optimum annealing temperature between $900\text{--}1000$\si{\degreeCelsius}, a maximum single-emitter fraction of \SI{18}{\percent} (\SI{11}{\percent}) and a single-emitter yield of $\eta = 0.09\,\%$ ($0.05\,\%$) for Bi$^{2+}$ (Sn$^{2+}$). 

To estimate the optimal dose, the Bi$^{2+}$ and Sn$^{2+}$ arrays in the \SI{900}{\degreeCelsius} annealed sample are further assessed. This is accomplished first by assuming a Poisson model describes the data and then analysing the count rates of single emitters for each array and identifying the least bright single emitter, which was then used as an adaptive threshold to count the number of spots with multiple emitters. In this way, the spots were divided into three groups: single emitters, multiple ($\geq2$) emitters, and spots where the implantation was unsuccessful; this includes the spots with fluorescence attributed to residual damage. Each array spot was treated as an independent trial, and a Poissonian model with mean number of emitters in each site $\lambda$ assumed to scale linearly with implant dose $d$ and formation probability in emitters per ion ($\alpha$) was used. Single emitter yield, $\eta$, and $\alpha$ are related to one another, with $\alpha$ being model-dependent but allowing us to take account of multiple emitters. The model was optimised to fit the recorded statistics by maximum likelihood estimation; the value of $\alpha$ that maximises the log-likelihood is denoted $\hat\alpha$ (see supplementary materials for more details).

The fitted model is shown in Fig.~\ref{fig:Poisson_analysis} along with the formation probabilities and the resulting $\hat\alpha $ values. The $P$-values from the fit were 0.24 (0.98)  for Bi$^{2+}$ (Sn$^{2+}$), confirming the goodness of the fit. The shaded regions around each fit indicate the standard deviation of the residuals of each fit.

\begin{figure}
        \includegraphics[width=1\linewidth]{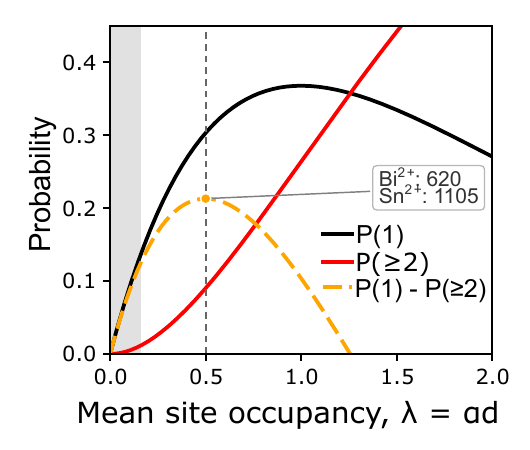}
        
    \caption{Modelled distributions (probability of different numbers of emitters per spot) for multiple and single emitters as a function of the mean site occupancy ($\lambda = \alpha d$). The difference distribution is also shown. The grey shaded region indicates the maximum extent of $\lambda$ found in this work ($\lambda = 0.16$ for Bi$^{2+}$). The peak in the difference distribution at $\lambda$ = 0.5 corresponds to doses of 620 and \SI{1105}{ions/spot} for Bi$^{2+}$ and Sn$^{2+}$, respectively.}
    \label{fig:Poisson_model}
\end{figure}

From the fitted $\hat\alpha$-values, an estimated distribution of multiples and singles at higher doses was modelled. These distributions are shown in Fig~\ref{fig:Poisson_model}. The distributions of single emitters are maximised at 1/$\hat\alpha$, and this occurs at $d=1235$ (\SI{2210}{ions/spot}) for Bi$^{2+}$ (Sn$^{2+}$). The difference distribution is also shown, and the peak of this distribution occurs at 1/2$\hat\alpha$ or 620 (\SI{1105}{ions/spot}) for Bi$^{2+}$ (Sn$^{2+}$). The peak in the difference distribution can be interpreted as the optimum dose for applications, where the difference between the probabilities for single and multiple emitters needs to be largest. The grey-shaded region indicates the maximum extent of the mean site occupancy covered in this paper.

\subsection*{Verification of emitter identity}

The existing set of measurements was extended to characterise the emitter type and propose a fast way to identify PL4-PL6. Four additional measurements were performed on each candidate emitter:  ODMR over \SI{1300}{\mega\hertz} - \SI{1400}{\mega\hertz} at zero magnetic field, low-temperature photoluminescence spectra (LTPL) at \SI{4}{\kelvin} over \SI{1030}{\nano\meter} - \SI{1140}{\nano\meter} with a \SI{80}{\giga\hertz} resolution spectrometer, polarisation detection of the emitted photons (POL) by projecting onto two orthogonal linear polarisations (H/V) (See methods below) and a saturation count rate measurement (SAT). Example spectra are provided in the supplementary information, and evaluation details are in the methods section. Table \ref{tab:emitter_identification} summarises the literature values used for identification \cite{shafizadeh2025}.

\begin{table*}[ht]

\centering

\begin{tabular}{c||c|c|c}
emitter type & ODMR frequency (MHz) & ZPL wavelength (nm) & polarisation dependence\\ \hline\hline
PL4 & 1334 & 1078 & \textit{no}\\ \hline
PL5 & 1344 \& 1375 & 1041.9 & \textit{yes}\\ \hline
PL6 & 1352 & 1037.7 & \textit{no}\\ 
\end{tabular}

\caption{Summary of centre frequency of room-temperature ODMR dips \cite{shafizadeh2025} and the position of zero phonon lines (ZPL) in low-temperature PL measurements to distinguish PL4-6 \cite{shafizadeh2025}.}

\label{tab:emitter_identification}

\end{table*}

\begin{figure*}[ht!]
    \includegraphics[width=\linewidth]{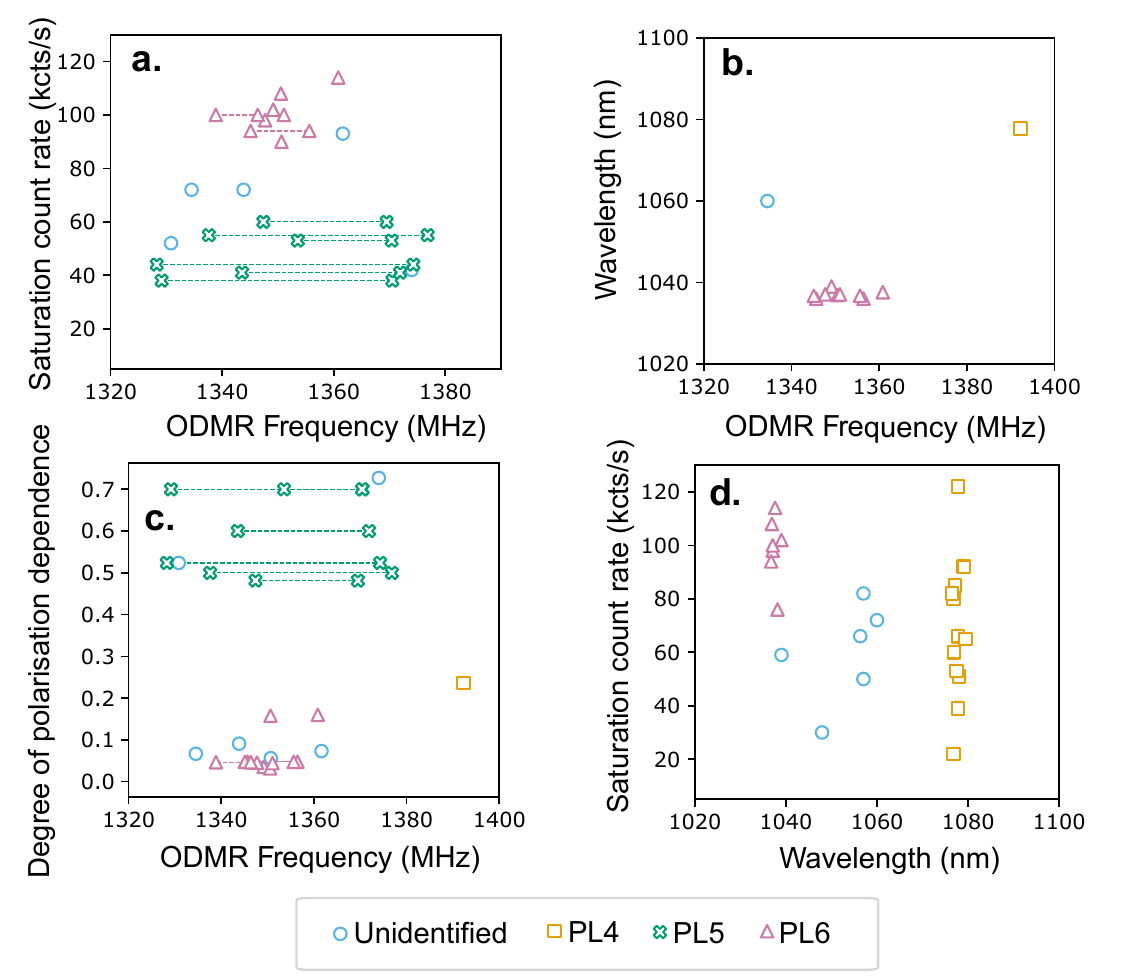}
    
    \caption{Pairwise scatter plots showing the correlations between different characterisation metrics. Each data point represents a single emitter in either the Bi$^{+/2+}$ or Sn$^{+/2+}$ in the \SI{900}{\degreeCelsius} sample. Blue circles denote unidentified emitters, yellow squares PL4, green crosses PL5 and pink triangles PL6. \textbf{a.} Saturation count rate as a function of ODMR frequency. PL6 and PL5 centres form distinct clusters. Dashed lines connect the two ODMR frequencies of emitters with split resonances. These indicate hyperfine coupling to a nearby nuclear spin for the PL6 centres. \textbf{b.} ZPL wavelength as a function of ODMR frequency. No PL5 ZPLs were observed during this study. \textbf{c.} Degree of polarisation dependence (DOP) with ODMR frequency. Clusters are visible for the PL5/PL6 centres. \textbf{d.} Saturation count rate as a function of ZPL wavelength. Both PL4 and PL6 are visible in this plot, with PL4 centres showing a larger spread in intensities.}
    \label{metric_correlations}
\end{figure*}

Across all dose arrays for Sn$^{2+}$ and Bi$^{2+}$ annealed at \SI{900}{\degreeCelsius}, 60 isolated single emitters exhibited distinct ODMR resonances and/or zero-phonon lines, which were evaluated using pairwise correlation plots (Fig \ref{metric_correlations}).

Clustering is observed in all plots, serving as unique fingerprints for different emitter species. Under the chosen measurement conditions, only the PL6 emitter consistently exhibit signals across all four metrics. The ODMR-SAT and ODMR-POL correlations imply that a simple thresholding at \SI{75}{kcps} (SAT) and \SI{0.3}{} (POL) allows for the separation of PL5 from PL6 (Figs. \ref{metric_correlations}a, c). These threshold values depend on the experimental setup and must be calibrated for each apparatus. Also, for the PL6 centre, no polarisation dependence is expected; the small observed degree of polarisation dependence (DOP) is attributed to polarisation-dependent losses and detection efficiency of the setup. 

Among the identified emitters, we observe a variation of the ODMR frequency. For PL6 centres a single ODMR dip is expected at zero magnetic field. A mean value of \SI{1351.7}{MHz} with a standard deviation of \SI{4.7}{MHz} is observed. To exclude misassignments, only emitters for which a single ODMR peak was observed were used. Split resonances for PL6 centres are likely caused by coupling to nearby nuclear spins. For the PL5 centres, two ODMR dips are expected at zero magnetic field. These values are reported separately for each dip. The PL5 dip 1 mean is \SI{1339.9}{MHz} with a standard deviation of \SI{10.1}{MHz}. For dip 2, the mean is \SI{1372.2}{MHz} and the standard deviation is \SI{2.8}{MHz}. This variation is attributed to strain induced by the heavy-ion implantation process, where significant local lattice damage is produced and may be present even after annealing. Additionally, the spread in ODMR frequencies is different for each dip in the PL5 spectra, suggesting that the two transitions respond differently to local strain. It has previously been shown that heavy-ion implantation damage is not fully healed at 800\text{--}1000\si{\degreeCelsius} annealing temperatures, with complete recovery reported only at substantially higher temperatures \cite{kobayashi2022}.

The zero-phonon lines in the LTPL spectra contribute mainly to the identification of PL4 centres. Figure \ref{metric_correlations}d shows that saturation count rate alone does not allow the distinction of PL4 from other emitter types. For the identification of PL6 centres, LTPL measurement is not necessary; the other metrics are sufficient, and LTPL can therefore be omitted, avoiding the need for a cryostat.  The measurement parameters used in this study do not permit detection of PL7 and PL8, which are also attributed to the oxygen vacancy. PL7 falls below the applied count rate threshold, and PL8 is not resolved due to its lower spin contrast and the chosen laser power and integration time \cite{hu2026}. Although PL4 centres are present in the LTPL spectra, the low spin contrast at room temperature limits the detection of PL4 ODMR signals, with only one being observed. 

Taken together, these correlations define a tiered characterisation procedure that minimises measurement cost to identify single emitters: 1) grid-fit the PL map and apply an intensity threshold to isolate bright single-emitter candidates, removing the majority of locations and most multiples. 2) Add a fast POL measurement to distinguish between PL5 and PL6. 3) Confirm the single-emitter character using $g^{(2)}(\tau)$ and the precise type using ODMR, reserving 4) LTPL measurements for cases where PL4 identification is required or where ODMR is inconclusive.

\section*{Discussion \& Conclusion}

We have demonstrated the maskless creation of arrays of bright, spin-active single emitters in 4H-SiC through low-energy FIB implantation of Bi$^{+/2+}$ and Sn$^{+/2+}$. Under optimal implantation and annealing conditions, up to 18$\%$ of implanted sites host a single emitter, with a single-emitter conversion yield of ~0.09$\%$ per ion. An implantation accuracy of \SI{146}{\nano\meter} is demonstrated for Bi$^{2+}$. Emitters are identified using a combination of ODMR, low-temperature PL, fluorescence saturation measurements and polarisation-dependent emitter emission. Correlations between these metrics define a strategy to rapidly characterise future samples. Together, these results establish maskless heavy-ion implantation as an efficient route to positioned single emitters in silicon carbide for quantum technologies.

An optimum annealing window of 900\text{--}\SI{1000}{\degreeCelsius} is demonstrated for all four species. Below \SI{900}{\degreeCelsius}, the implantation-induced vacancies lack the mobility required to migrate and form stable, optically active complexes, while at annealing temperatures of \SI{1100}{\degreeCelsius} the arrays are almost completely lost, suggesting that at elevated temperatures the thermal energy drives emitter annihilation or aggregation into larger, non-optically-active clusters. This is in line with emitter evolution reported for heavy-ion-implanted 4H-SiC, in which silicon vacancies convert to divacancies and related complexes around \SI{1000}{\degreeCelsius}, and transform into TS emitters and other clusters at higher temperatures \cite{kobayashi2022}. The window is comparable to that reported for formation with lighter ions \cite{lee2021}, indicating that the damage caused by heavy ions does not qualitatively alter the thermally activated formation pathway. The fluorescence observed at filtered-out array locations is attributed to residual implantation damage, and among the identified PL5 and PL6 centres we observe substantial scatter of the ODMR frequencies (standard deviations of 4.7 MHz for PL6 and up to 10 MHz for PL5). This is attributed to local strain persisting after annealing, a consequence of the dense collision cascades.
    
Across all dose arrays for  Bi$^{2+}$ and  Sn$^{2+}$ annealed at \SI{900}{\degreeCelsius}, 60 isolated single emitters were characterised and exhibited a distinct ODMR resonance and/or low-temperature ZPL. In the Bi$^{2+}$ \SI{200}{ions/spot} array, which yielded the most single emitters, 5 of the 18 $g^{(2)}$-confirmed single emitters were identified as PL6 centres and 2 as PL5 centres. In the corresponding Sn$^{2+}$ array, 1 of 6 single emitters was identified as PL6, while no PL5 centres were identified. 

Formation of PL5 and PL6 centres has previously been demonstrated using focused helium ion beam (He-FIB) implantation and masked carbon implantation. In the He-FIB study, a 10x10 array implanted with a dose of 300 ions/spot showed ODMR and low-temperature PL spectra in 98 $\%$ locations; however, PL6 centres were located in 5 locations, and only 1 was designated as a single emitter. PL5 centres were not recorded in this array \cite{he2024}. This places the single emitter yield ($\sim$ 0.007 $\%$) and the total number of recorded single emitters below those demonstrated in this study. Masked carbon implantation reached a higher per-ion yield of $\sim$0.7$\,\%$ at \SI{20}{ions/spot}, owing to the much lower dose, with an overall single-emitter site fraction of $\sim$14$\,\%$ and per-site fractions of $7\,\%$ PL5 and $1\,\%$ PL6 \cite{li2022}. Given the recent assignment of PL5-8 to oxygen-vacancy complexes, it is unsurprising that the species-specific yields remain at the few percent level for all approaches in which no oxygen is supplied. The formation of PL5 and PL6 centres in the absence of implanted oxygen ions is attributed to the residual lattice or surface-bound oxygen that is incorporated during implantation and annealing. 

The doses in this study were deliberately restricted to the low-dose regime. At \SI{200}{ions/spot}, the mean site occupancy reaches only $\lambda = 0.16$ (Bi$^{2+}$) and $0.09$ (Sn$^{2+}$), so multiply-occupied sites remain rare. Reaching higher single-emitter numbers requires larger doses under the same annealing conditions, at the cost of an increasing fraction of multiply-occupied sites. From the distributions for Bi$^{2+}$ and Sn$^{2+}$ at \SI{900}{\degreeCelsius}, the inverse of $\alpha$ gives the dose for which $p_{1}$ is maximised, and this occurs at 1235 and \SI{2210}{ions/spot}, respectively, for  Bi$^{2+}$ and Sn$^{2+}$. Estimations for the optimum dose can be gained from these distributions; it can be seen that the value of $\lambda$ that maximised the percentage of single emitters while minimising multiples occurs at $1/2\alpha$  (620 and \SI{1100}{ions/spot}). 
      
For future sample characterisation and device development, a rapid characterisation procedure for PL6 centres that reaches a confident emitter-type assignment can be achieved through grid fitting and intensity thresholding to isolate bright single-emitter candidates, complemented by a fast polarisation measurement, $g^{(2)}(\tau)$ confirmation of single-photon character and ODMR for type assignment. Low-temperature spectroscopy, the most time-consuming characterisation step, did not prove necessary for PL6 identification and is required mainly where PL4 assignment is sought. The assessment of correlations between these metrics shows clear clustering for PL5 and PL6 centres when considering both saturation count rates and the degree of polarisation as a function of ODMR frequency.

Future studies will address the species-specific yield directly. Maskless array-generation with oxygen-containing gas species, delivered through the Ionoptika QOne duoplasmatron source, should increase the generation yield of PL5 and PL6 centres, while retaining precise control over their number and position. Combined with the intensity-threshold screening and correlated measurement scheme introduced here, this points towards the high-efficiency formation and rapid identification of the large numbers of PL6 centres needed to advance scalable silicon carbide quantum technologies.

\section*{Methods}

\subsection*{Substrates}

4H-SiC samples were fabricated and provided by the IISB in accordance with LIST's specifications: 10.74 $\mu$m epilayer doped with $1.18\times10^{15}$\,{cm}$^{-3} $ N on a commercially available n-type substrate wafer.

\subsection*{Implantation}

Maskless implantation patterning was performed at the UK National Ion Beam Centre using a FIB (Ionoptika, QOne) running a eutectic BiSn liquid metal alloy ion source. A Wien (\textbf{E}$\times$\textbf{B}) filter selects the specific mass-to-charge state of the species, allowing choice of either Bi or Sn in either singly or doubly ionised states. This Wien filter can be narrowed to allow selection of a specific isotope, and in this study we used the $^{120}$Sn and $^{209}$Bi isotopes. 

An anode voltage of \SI{25}{kV} was used for all implants in this study, and therefore doubly charged ions arrive at the surface with an energy of \SI{50}{keV}. 

The depth profiles for the ions was estimated using a Monte Carlo model (SRIM\cite{ziegler2010}). The implant is  peaked at \SI{14.7}{\nano\meter}/\SI{15.4}{\nano\meter} (\SI{25}{keV} Bi/Sn) and \SI{21.0}{\nano\meter}/\SI{23.8}{\nano\meter} (\SI{50}{keV} Bi/Sn) and the straggle varies between 2.6\text{--}\SI{5.2}{\nano\meter}. Further details regarding sample and implantation conditions are provided in the supplementary materials.

The FIB system allows maskless control over the spatial position of the implant with \SI{14}{\nano\meter} resolution (assessed by the feature size in  FIB imaging mode with Sn$^{2+}$, shown in the supplementary materials). 
An electrostatic chicane removes neutral atoms that would otherwise be unaffected by beam steering and blanking and deposited in an uncontrolled way. 

The incident ion current was measured using a secondary-electron-suppressed Faraday cup adjacent to the samples immediately before implantation. This current was monitored during patterning using upstream apertures in the column and measured again after implantation \cite{schneider2026, cassidy2021}. Typically, the change in beam current before and after implantation was less than 2$\%$. 

To write complex patterns using this small beam spot, Nabity Nanoscale Pattern-Generation Software was used in combination with electrostatic scanning plates and a fast electrostatic blanker with a 10 ns rise time.

The pulse width of the blanker was adjusted to deliver a target number of ions, with an uncertainty governed by Poisson statistics and the stability of the beam current given above. 

In total, nine identical 4H-SiC samples with epilayers were implanted. The primary experimental patterns were arrays of $10\times10$ resolution-limited spots of constant species and dose. Doses ranged from 10~ions/spot to 200~ions/spot.

\subsection*{Annealing and cleaning}

Eight of the nine samples were annealed in a tube furnace under an argon atmosphere at temperatures ranging from \SIrange{600}{1100}{\degreeCelsius}, and one sample was retained as an unannealed reference. A 2-hour dwell time with a ramp rate of \SI{5}{\degreeCelsius\per\minute} was used. Before optical measurements, samples were cleaned in an IPA ultrasonic bath for \SI{20}{\minute} and subsequently immersed in Piranha acid for \SI{15}{\minute} with a 3:1 ratio of concentrated sulfuric acid and \SI{30}{\percent} hydrogen peroxide.

\subsection*{Optical measurements}

Photoluminescence mapping and Hanbury-Brown and Twiss (HBT) measurements were performed at room temperature using a purpose-built confocal microscope with an off-resonant \SI{940}{\nano\meter} laser (Roithner RLTMDL-940L-200-2S). The wavelength was chosen to suppress excitation and therefore the competing emission of the silicon-vacancy V$_\text{Si}^{-}$. The microscope is composed of a 0.9 NA near-infrared objective (Zeiss Epiplan Neofluar) mounted in a piezo scanner (Mad City Labs Nano LP 300). The emission is separated from the laser and filtered by a \SI{950}{\nano\meter} long-pass dichroic mirror (Thorlabs DMLP950), a \SI{1000}{\nano\meter} long-pass cut-on filter (Thorlabs FELH1000) and a \SI{1326}{\nano\meter} short-pass filter (Semrock) before coupling into a 1060XP optical fibre. These filters were used to define a spectral window that covers the room-temperature phonon sideband emission of PL1-8 centres fully, while suppressing V$_\text{Si}^{-}$ emission. A data aquisition card (National Instruments USB-6363) and a TimeTagger Ultra (Swabian Instruments) together with NIR superconducting nanowire single-photon detectors (SNSPDs) (IDQuantique ID281) were used to record the data.

The normalised $g^{(2)}(\tau)$ for each HBT measurement was fitted using $ g^{(2)}(\tau) = ((N-1)/N) + (1/N)(1-(1+a)e^{-|\tau|/\tau_{1}} + ae^{-|\tau|/\tau_{2}})$ where $\tau$ is the time delay between detection events, the decay constants $\tau_{1,2}$ are related to the transition rates between energy levels in the emitter system, $a$ accounts for the presence of decay channels with meta-stable states, and the number of emitters is $N$. Throughout all the measurements, no background correction was applied.

Count rate measurements as a function of excitation power were used to extract the saturation intensity. Details of the fitting process are given in the supplementary materials. 

The dependence of the count rate on excitation polarisation was also assessed by measuring the maximum and minimum count rates as the polarisation of the excitation beam was varied. The degree of polarisation dependence (DOP) was then calculated using $ (I_\text{max} - I_\text{min}) / (I_\text{max} + I_\text{min})$. A DOP of 1 indicates a high degree of dependence on input polarisation, while a DOP of 0 indicates no preference on excitation polarisation. 

\subsection*{cw-ODMR}

Microwave signals were generated using an amplified signal generator (Rohde $\&$ Schwarz SMIQ 03B + Mini-Circuits ZHL-25W-272+) before being passed to a custom-designed microwave antenna fixed to the objective above the sample.

We extracted the centre frequency of each ODMR dip by fitting a Lorentzian to each dip in the ODMR spectra 
\[
I(f) = I_0 - \sum_{i=1}^{M} A_i \frac{(\Delta f_i / 2)^2}{(f - f_{0,i})^2 + (\Delta f_i / 2)^2}
\]
where $I(f)$ represents the normalised ODMR contrast at a given applied microwave frequency $f$, and $I_0$ denotes the baseline signal, $A_i$ is the amplitude of the dip, $f_{0,i}$ denotes the centre frequency, and $\Delta f_i$ represents the full-width at half-maximum (FWHM) linewidth. The summation accommodates multiple distinct spin resonance transitions and therefore accounts for the double dip in PL5 spectra ($M=2$), as well as the single resonance of PL6 ODMR ($M=1$).

\subsection*{Low Temperature Spectra}

Low-temperature measurements were performed using a confocal setup with the sample mounted in a $\SI{4}{\kelvin}$ cryostat (Attocube, attoDRY800). Spectra were recorded on a spectrometer (Teledyne HRS-750 with NIRvana HS camera) under \SI{910}{\nano\meter} laser excitation (Sirah Matisse CR). As with the choice of \SI{940}{\nano\meter} excitation, \SI{910}{\nano\meter} also suppressed the excitation of the V$_\text{Si}^{-}$ centres.

\subsection*{Data Availability}

The datasets generated and/or analysed during the current study are available in the "Data for: Efficient formation and identification of single emitters in 4H-SiC following maskless heavy ion implantation" repository, DOI: 10.5281/zenodo.22125467.


\newpage
\onecolumn
\bibliographystyle{naturemag}
\bibliography{bibliography_zotero}


\section*{Funding}
 
This work was supported by the Surrey Ion Beam Centre, which is funded by UK EPSRC (Grant No. EP/X015491/1). SE, OGLW, and ML are grateful to EPSRC for their PhD studentships.

L.K.S.Z, S.K., and F.K. acknowledge support by the Luxembourg National Research Fund (FNR) via the PEARL chair "AQuaTSiC" under grant agreement 17792569. F.K. additionally acknowledges the FNR for the project "SiCqurTech" under the national grant
agreement 18253399, as well as the European Union’s Horizon 2020 Research and Innovation Programme under grant agreement
101017733. F.K. received additional support via the European Research Council for the project "Q-Chip" under
grant agreement 101171067, and the Horizon Europe Programme for the Flagship project "QIA Phase 1" under grant agreement 101102140.

\section*{Author contributions}

\textbf{SE, SKC, BNM, LKSZ, SK, FK} devised the project, \textbf{MH, PB} provided substrates, \textbf{ES, ML, KS, GA, LA, SE} performed ion implantation, annealing and related analysis, \textbf{ES, LKSZ, JH, SK} performed PL and related experiments, \textbf{SE, LKSZ, OGLW, SK} performed data analysis, \textbf{SE, BNM, LKSZ, SK, FK} wrote the manuscript, and all authors reviewed the manuscript.

\section*{Competing interests}

All authors declare no financial or non-financial competing interests. 


\clearpage
\onecolumn

\begin{center}
{\Huge\bfseries Supplementary Information\par}
\vspace{1cm}
\end{center}

\setcounter{section}{0}
\setcounter{equation}{0}
\setcounter{figure}{0}
\setcounter{table}{0}

\renewcommand{\thesection}{S\arabic{section}}
\renewcommand{\theequation}{S\arabic{equation}}
\renewcommand{\thefigure}{S\arabic{figure}}
\renewcommand{\thetable}{S\arabic{table}}

\section{Ion implantation}

\subsection{Implantation depth and pattern}

Ion implantation patterning was performed using the Ionoptika Q-One Single Ion Multispecies Positioning at Low Energy (SIMPLE) tool at the UK National Ion Beam Centre. SIMPLE is a focused ion beam (FIB) system that extracts a high-brightness beam from a liquid metal alloy ion source (LMAIS). Ions are generated via field ionisation at a sharp liquid-metal-coated tip and accelerated through 10–25 kV. In this work, a BiSn eutectic alloy was used, enabling rapid switching between Sn and Bi species as well as singly and doubly charged states within a single run.

\begin{figure}[!ht]
    \centering
    \includegraphics[width=0.6\textwidth]{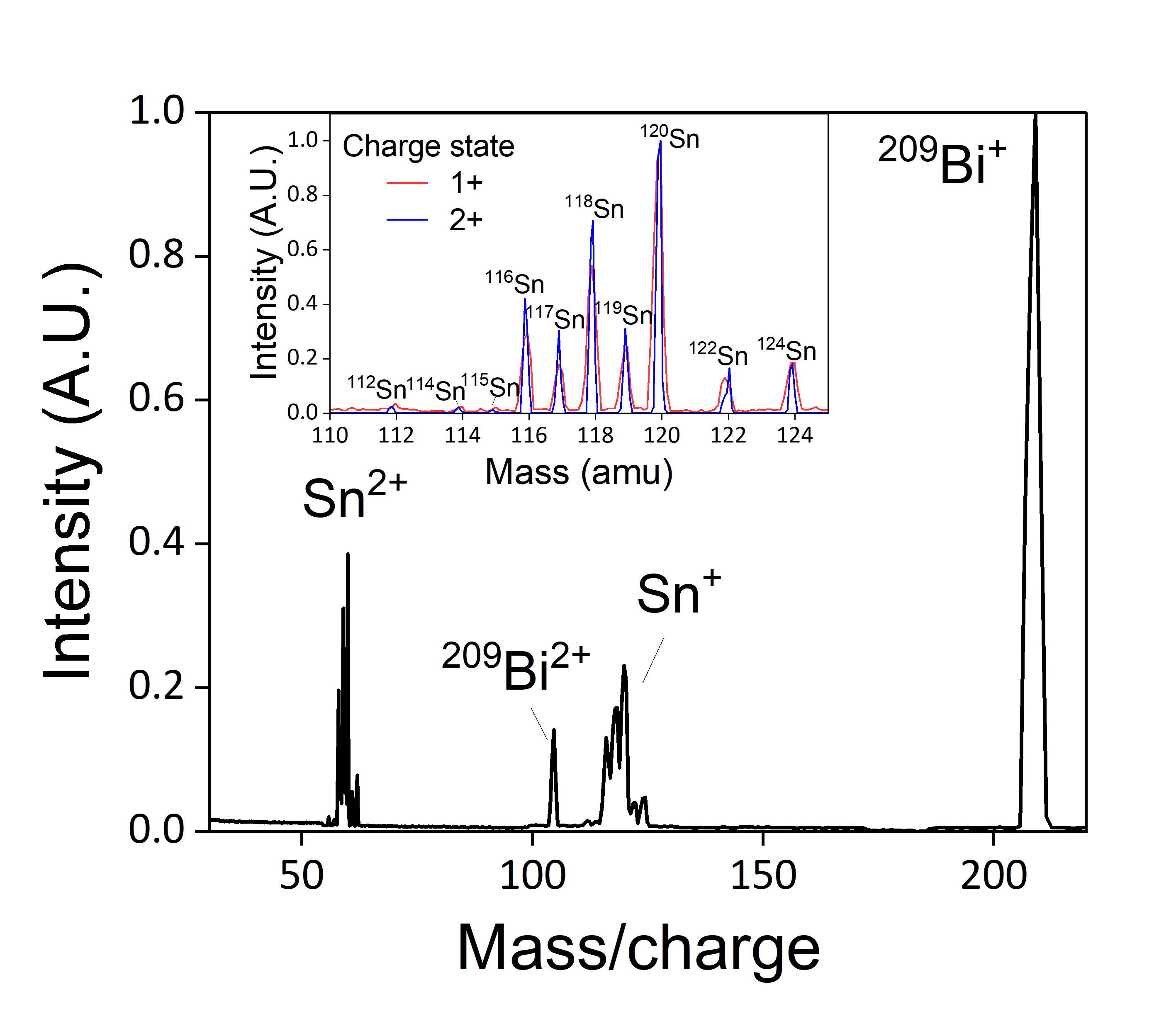}
    \caption{Wien filter mass/charge spectrum of the BiSn liquid metal alloy source taken at a 25kV acceleration voltage with a 70$\mu m$ beam-defining aperture. The inset shows the singly and doubly charged tin isotopes taken with a 10$\mu m$ aperture to achieve higher mass resolution.}
\label{fig:BiSn_WFscan_NPGS}
\end{figure}

A Wien filter ($E\times B$) selects the specific mass-to-charge ratio ($M/q$) of interest, while neutral atoms are deflected using an electrostatic chicane filter. Mass spectra (shown in Figure \ref{fig:BiSn_WFscan_NPGS}) were recorded by sweeping the Wien filter potential and measuring current at a Faraday cup on the sample stage. To achieve clear separation of Sn isotopes, the system was operated in high mass resolution mode by condensing the beam into the $\mathbf E \times \mathbf B$ filter alongside a $10 ~\mu \text{m}$  resolving aperture. The $^{120}$Sn isotope was selected for Sn implants, while Bi features a single isotope ($^{209}$Bi). Mass resolution values ($M/\Delta M$), derived from Gaussian fits, were determined as $755 \pm 5$ for $^{120}\text{Sn}^{2+}$, $460 \pm 33$ for $^{120}\text{Sn}^{+}$, $104 \pm 20$ for $^{209}\text{Bi} ^+$, and $364 \pm 15$ for $^{209}\text{Bi}^{2+}$.

\begin{figure}[h!]
    \centering
    \includegraphics[width=0.7\linewidth]{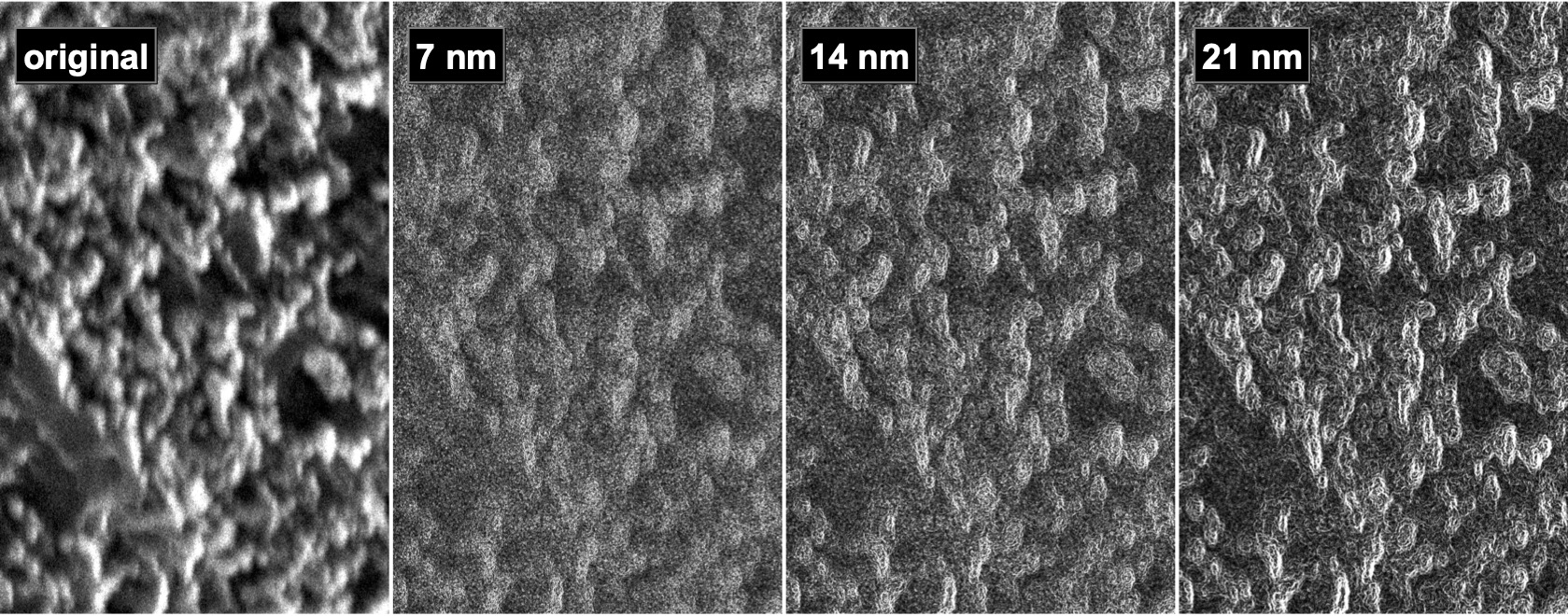}
    \caption{Imaging resolution of 50keV Sn$^{2+}$ beam. At left: a secondary electron image of texture on Faraday cup (12 $\mu$m along the base). Gaussian gradient filters of successively larger s.d. have then been applied (each result has been independently contrast-corrected). The 7~nm  filtered image shows almost no clear, continuous edges; at 14~nm many features have edges, and at 21~nm most features have a clear edge. We take the imaging resolution to be on the order of 14~nm. }
    \label{fig:ImageRes}
\end{figure}

Downstream electrostatic lenses, plates, and apertures control beam focusing, blanking, scanning, and collimation. The 10~$\mu$m  collimator aided in minimising beam spot size, which was evaluated by imaging substrates with topographical features smaller than the beam profile. The sharpest profile edges provide an upper bound for the focused beam diameter, demonstrated in Figure \ref{fig:ImageRes} with a beam diameter of order 14~nm.

The beam current was measured using a secondary-electron-suppressed Faraday cup connected to a Keithley picoammeter. Current values were calculated from the mean difference between blanked and unblanked states over a 10 s integration period (repeated twice). Operating currents ranged from 0.3 to 5 pA and were verified before and after patterning, typically agreeing within 2\%. Continuous upstream monitoring through a column aperture confirmed beam current stability remained within 1\% throughout implantation.

Precise control over the ion dose was achieved down to single-ion precision using a fast electrostatic blanker with a 10~ns rise time. Implants were performed in a Poissonian mode, where the pulse width was adjusted to deliver the required average number of ions per spot (with the variance governed by Poisson statistics). For example, at a typical beam current of 2.1~pA, a 700~ns pulse yields an average dose of 10 ions/spot for singly charged species.

\begin{figure}[h!]
    \centering
    \includegraphics[width=0.45\linewidth]{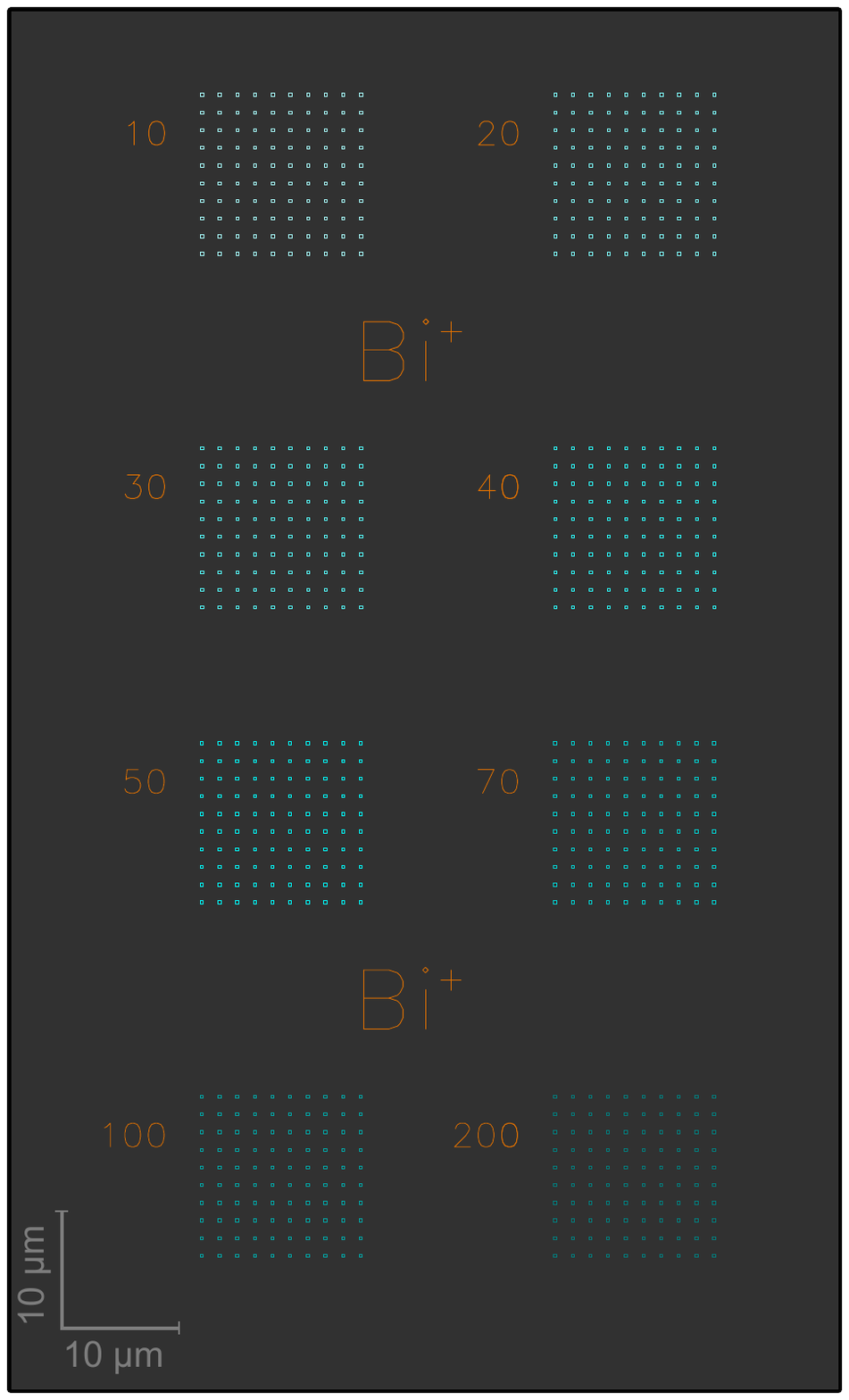}
    \caption{Example of an implant pattern. Orange writing indicates the directly written label for implanted species and dose in ions/spot. Blue squares show the 10x10 point array for each dose.}
    \label{fig:Nabity}
\end{figure}

\begin{figure}[h!]
    \centering
    \includegraphics[width=\linewidth]{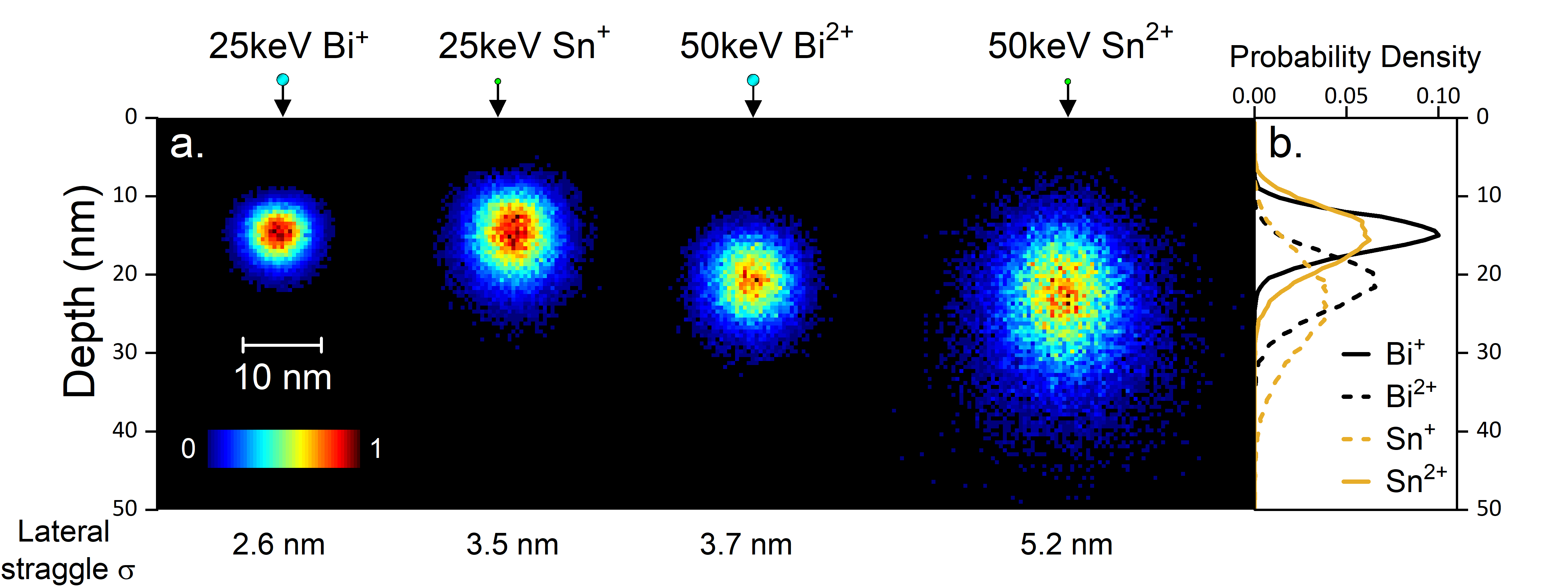}
    \caption{SRIM simulations of the Bi$^{+/2+}$ and Sn$^{+/2+}$ implant distributions in a SiC substrate for an anode voltage of 25~kV. a) shows 2D profiles of the normalised frequency density against the lateral distance and depth, and b) shows 1D line profiles of the probability density as a function of depth.}
    \label{fig:2DTRIM}
\end{figure}

Spatial positioning was controlled through a dedicated Nanometer pattern generator connected to SIMPLE. The point arrays were written according to the layout in Figure \ref{fig:Nabity}, consisting of four $10\times10$ point arrays with a $3~ \mu $m pitch. Markers specifying species and ion dose per spot were written at a fluence of $10^{15}$ ions cm$^{-2}$ to ensure bright photoluminescence emission. To prevent unwanted exposure, the beam was electrostatically blanked during movement between array sites. Implantation was carried out across a $120 \times 120 ~\mu \text{m}^2$ write-field per window, with samples mounted at a 3$^\circ$ tilt off-axis to suppress ion channelling. The high-vacuum sample chamber ($10^{-8}$~mbar) ensured the samples were not contaminated during the implantation process. Monte Carlo SRIM simulations (Figure \ref{fig:2DTRIM}) estimate projected implantation depths of 10–40 nm and lateral straggle of 2.6–5.2 nm across the implanted species.

\subsection{Positional accuracy of implantation}

The precision of the implantation was quantified by assessing the variance in the position of the emitters compared to a simulated grid. The grid was fitted to the photoluminescence map by allowing for rotation and lateral translation and minimising the squared distance for the whole grid simultaneously. From the simulated grid, Gaussian fittings are then applied to each identified emitter, and the root mean squared difference (RMS) between the peak position of each Gaussian fit and the corresponding simulated grid position is calculated and shown in Fig. 1b. of the main text. This RMS difference has previously been shown to be dominated by the implantation spot, with minimal contribution from annealing and implantation straggle, therefore, this provides a measure of implantation accuracy and is shown as a red circle in Fig. 1b. with a value of 146~nm for the Bi$^{2+}$ 200 ions/spot array in the sample annealed at 900$^\circ $C. 

The RMS displacement is calculated using
$\sigma_\text{RMS} = \sqrt{(1/N) \Sigma_{i} [(x_{i} - \hat{x}_{i})^{2} + (y_{i} - \hat{y}_{i})^{2}]}$.  where ($x_{i}$, $y_{i}$) is the Gaussian-fitted centroid of emitter i, ($\hat{x}_{i}$, $\hat{y}_{i}$) the nearest node of the best-fit ideal grid, and N the number of emitters retained by the fitting.

\section{Optimal dose estimation}

To estimate the implant dose that optimises single-emitter formation, we model the site-occupation statistics using a Poisson model. The mean site occupancy, $\lambda$, is assumed to scale linearly with dose in the low-dose regime, i.e. $\lambda=\alpha d$, where $d$ is the implant dose in ions/spot and $\alpha$ a constant of proportionality (the formation probability in emitters per ion).  Each array position was treated as an independent trial that may contain zero, one, or multiple bright emitters, with the probability of finding $k$ emitters at a site given by the Poisson distribution $p_k(d) = (\alpha d)^{k}e^{-\alpha d}/k!$.
Sites are classified as empty ($k=0$), single ($k=1$), or multiple ($k\ge2$), giving respective probabilities
\begin{align*}
    p_0(d) &=  e^{-\alpha d} ,\\
    p_1(d) &=  \alpha d e^{-\alpha d} ,\\
    p_{\ge2}(d)  &= 1 - p_0(d) - p_1(d) = 1 - e^{-\alpha d} - \alpha d e^{-\alpha d}.
\end{align*}

The most probable $\alpha$ can be extracted from the experimental data through Maximum-Likelihood-Estimation. At each dose $d_i$, $N$ implanted spots are measured. From this data the number of spots containing exactly one emitter ($n_{1,i}$) and two or more emitters ($n_{\ge2,i}$) can be obtained, the experimental data is shown in Fig. 3 of the main text, for Bi$^{2+}$ and Sn$^{2+}$. The number of "empty" spots is calculated as $n_{0,i} = N - n_{1,i} - n_{\ge2,i}$. The likelihood, $\mathcal{L}$, of observing the entire dataset across all doses, given a specific yield parameter $\alpha$, is the product of the probabilities 

\[
    \mathcal{L}(\alpha) = \prod_{i} \frac{N!}{n_{0,i}! n_{1,i}! n_{\ge2,i}!} [p_0(d_i)]^{n_{0,i}} [p_1(d_i)]^{n_{1,i}} [p_{\ge2}(d_i)]^{n_{\ge2,i}},
\]
where, to simplify computation and avoid numerical underflow, the natural logarithm of the likelihood function is taken. Since ${N!}/{n_{0,i}! n_{1,i}! n_{\ge2,i}!}$ is constant with respect to $\alpha$, it is ignored for the purpose of optimisation, leaving the log-likelihood function

\[
    \ln \mathcal{L}(\alpha) \propto \sum_{i} \Big( n_{0,i} \ln[p_0(d_i)] + n_{1,i} \ln[p_1(d_i)] + n_{\ge2,i} \ln[p_{\ge2}(d_i)] \Big).
\]
The optimal parameter, $\hat{\alpha}$, is the value that maximises this log-likelihood.

To quantify how well the fitted model represents the experimental data (as in the examples shown in Fig. 3), Pearson’s chi-squared ($\chi^2$) statistic is calculated. This test compares the observed counts ($O$) to the expected counts ($E$) predicted by the model at $\hat{\alpha}$. For each dose category $i$ and outcome state $j \in \{0, 1, \ge2\}$, the expected counts are $E_{i,j} = N \cdot p_j(d_i)$. The $\chi^2$ statistic is formulated as:$$\chi^2 = \sum_{i} \sum_{j \in \{0, 1, \ge 2\}} \frac{(O_{i,j} - E_{i,j})^2}{E_{i,j}}.$$ The number of degrees of freedom (DOF) for this test is determined by the number of independent data bins minus the number of fitted parameters. For each dose, there are three categories, but since they must sum to $N$, there are two independent bins per dose. With one fitted parameter ($\alpha$), the number of degrees of freedom is $ 2N_\text{doses} - 1$. The resulting $\chi^2$ value and DOF are used to calculate a corresponding $P$-value (the probability of observing the data given that the model and its $\alpha$ is correct) from the cumulative chi-squared distribution. A $P$-value greater $\ge 0.05$ indicates that the null hypothesis cannot be rejected and the model is satisfactory. As mentioned in the main text $P$ was always larger than 0.2.

The optimum dose can then be calculated from  $\hat\alpha$. The probability of a single emitter $p_1(d)$ is maximized for
$$d_\text{opt} = 1/\hat \alpha. $$ 
On the other hand, maximizing  the difference distribution $p_1(d) - p_{\ge2}(d)$ occurs when 
$$d_\text{opt} = 1/2\hat\alpha. $$
The latter might be desirable if separating multiple emitters from singles (by $g^{(2)}$ measurement) is costly and must be minimized, or if spots that are empty can be later corrected by a second implantation attempt, but spots where multiple emitters have appeared cannot.

\section{Example data}

\begin{figure}
    \centering
    \begin{subfigure}[b]{0.48\textwidth}
    \raggedright \textbf{a.} \par 
        \centering
        \includegraphics[width=\linewidth]{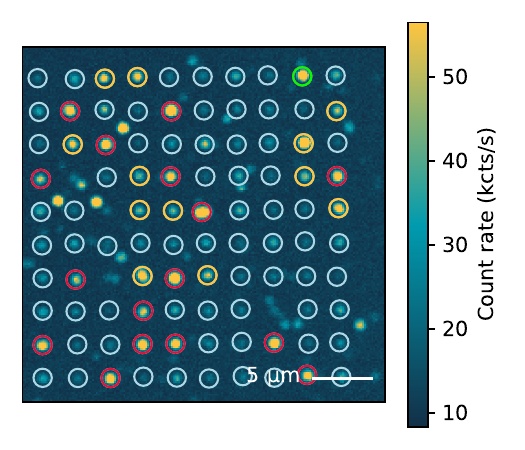}
        
    \end{subfigure}%
    \begin{subfigure}[b]{0.48\textwidth}
    \raggedright \textbf{b.} \par 
        \centering
        \includegraphics[width=\linewidth]{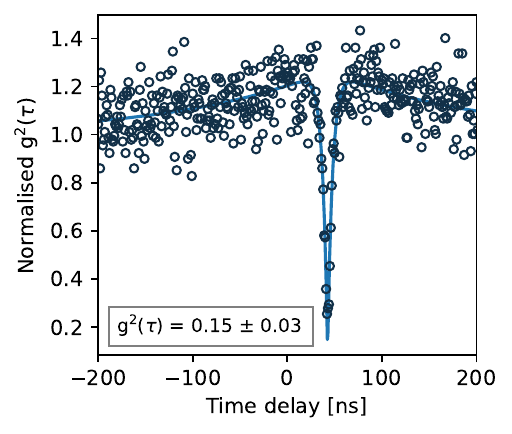}
        
    \end{subfigure}%
    \caption{\textbf{a.} The same confocal map as shown in Fig. 1a  of the main text. \textbf{b.} A typical HBT measurement for a single emitter, shown for the green circled emitter, with a characteristic antibunching profile for the second-order correlation function with $g^{2}(\tau) = 0.148\pm 0.029$.}
    \label{fig:pl_map_and_hbt_si}
\end{figure}

An example g$^{(2)}\tau$ measurement for the single PL6 centre circled in green in Fig. \ref{fig:pl_map_and_hbt_si} a, is shown in Fig. \ref{fig:pl_map_and_hbt_si} b, showing a characteristic dip in the second order correlation function  g$^{(2)}(\tau) < 0.5$ indicating a single emitter is present at this implant location.

\begin{figure}[ht!]
    \centering
 
    \begin{subfigure}[b]{0.48\textwidth}
    \raggedright \textbf{a.} \par 
        \centering
        \includegraphics[width=\linewidth]{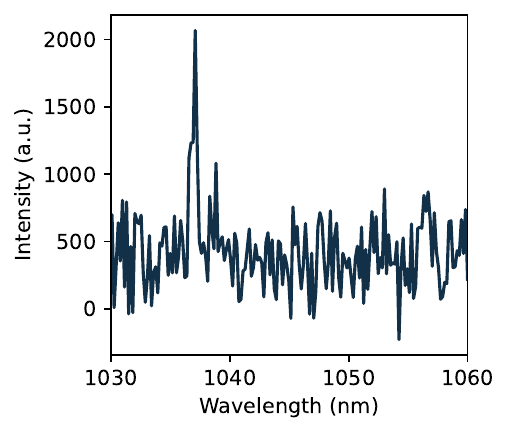}
    \end{subfigure}%
    \hfill  
    \begin{subfigure}[b]{0.48\textwidth}
    \raggedright \textbf{b.} \par
        \centering
        \includegraphics[width=\linewidth]{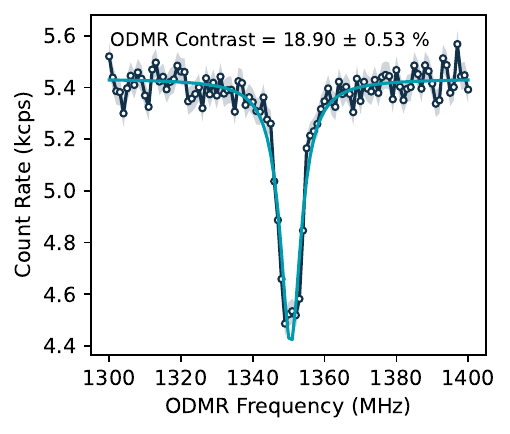}
    \end{subfigure}%
    \vspace{1.5em} 
  
    \begin{subfigure}[b]{0.48\textwidth}
    \raggedright \textbf{c.} \par
        \centering
        \includegraphics[width=\linewidth]{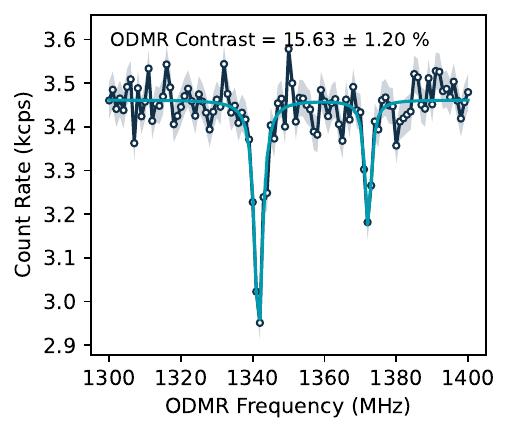}
    \end{subfigure}%
    \hfill   
    \begin{subfigure}[b]{0.48\textwidth}
    \raggedright \textbf{d.} \par
        \centering
        \includegraphics[width=\linewidth]{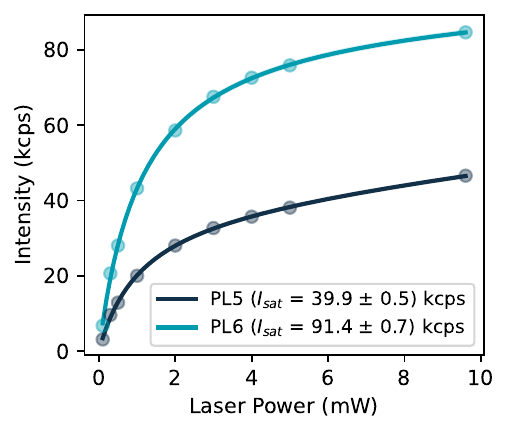}
    \end{subfigure}%

    \caption{\textbf{a.} Low temperature PL spectrum of a single PL6 centre with ZPL at  1038 nm. \textbf{b.} Room temperature ODMR spectrum of the same PL6 centre showing characteristic dip at 1350 MHz with a $\sim$ 7 MHz linewidth. \textbf{c.} Room temperature ODMR of a single PL5 centre showing typical double dip with centre frequencies of 1342 MHz and  1372 MHz with linewidths of 2.7 and  2.2 MHz, respectively and \textbf{d.} A comparison of the count rate saturation behaviour for one PL5 centre and one PL6 centre in the Bi$^{2+}$ 200 ions/spot array shown previously. The PL6 centre is the same green circled emitter as before and shows a saturation intensity of 91.4 $\pm$ 0.7 kcps, while the PL5 centre shows saturation count rates of 39.9 $\pm$ 0.5 kcps.}
    \label{fig:ODMR_LTPL_si}
\end{figure}

An example ODMR measurement for an identified PL6 centre is shown in Fig. \ref{fig:ODMR_LTPL_si} b under $100 \mu$W of optical excitation and 33 dBm of microwave power, the low temperature PL spectrum for the same PL6 centre is shown in Fig. \ref{fig:ODMR_LTPL_si} a, and characteristic ODMR and a PL5 centre in Fig. \ref{fig:ODMR_LTPL_si} c.

For the case of the PL6 centre, the single dip corresponds to the spin transition from $|0\rangle$ with m$_{s}$ = 0 to $|-1\rangle$ with ms = $\pm$1 in the absence of an external magnetic field. We can see from Fig \ref{fig:ODMR_LTPL_si} b, the single ODMR dip present for the isolated single emitter circled in green in Fig \ref{fig:pl_map_and_hbt_si} a. This ODMR spectra confirms its identity as a single PL6 centre with an ODMR centre frequency of 1350.54 $\pm$ 0.12 MHz, FWHM of 6.93 $\pm$ 0.37 MHz and ODMR contrast of 18.95 $\pm$ 0.65 $\%$. The ODMR spectra of a single PL5 centre is also shown in Fig \ref{fig:ODMR_LTPL_si} c with characteristic double resonance structure showing peaks at 1341.63 $\pm$ 0.10 MHz and 1372.16 $\pm$ 0.18 MHz with FWHMs of 2.67 $\pm$ 0.31 MHz and 2.24 $\pm$ 0.47 MHz, and ODMR contrasts of 15.64 $\pm$ 1.20 $\%$ and 8.43 $\pm$ 1.22 $\%$, respectively. No ZPLs were observed for the PL5 centres in this study.

Intensity saturation measurements are made for a selection of emitters. Two example measurements are shown in Fig. \ref{fig:ODMR_LTPL_si} d, for a single PL5 and single PL6 centre. Data are fitted using the steady-state saturation model

\begin{equation}
    I(P) = I_{\text{sat}} \frac{P}{P + P_{\text{sat}}} + c \cdot P
\end{equation}
where $I_{\text{sat}}$ represents the saturation single-photon emission rate, and $P_{\text{sat}}$ denotes the saturation excitation power. The linear coefficient $c$ is introduced to decouple the true emitter emission from background noise that scales linearly with the excitation laser power.

For an emitter under Poissonian (shot-noise-limited) statistics, the error in the count rate, $R$, with an integration time, $T$, is given by $\sigma_{R} = \sqrt{R/T}$. In Fig. \ref{fig:ODMR_LTPL_si} d, the error bars are too small to see. The same approach is used to define the error bars on data in Fig \ref{fig:ODMR_LTPL_si} b, c.


\end{document}